\documentclass[aps,twocolumn,amsmath,amssymb,footinbib,superscriptaddress,notitlepage]{revtex4-2}

\usepackage{graphicx}
\usepackage{amsmath,amssymb}
\usepackage{dcolumn} 
\usepackage{bm} 
\usepackage[utf8]{inputenc}
\usepackage{listings}
\usepackage{float}
\usepackage{epstopdf}
\usepackage{hyperref}
\usepackage[all]{hypcap}
\usepackage{natbib}
\usepackage{soul}
\usepackage{ulem}

\begin{document}

\title[Strain in 2D TMDCs induced by metal-assisted exfoliation from the polyvinilalcohol-covered substrate]{Strain in 2D TMDCs induced by metal-assisted exfoliation from the polyvinilalcohol-covered substrate}
\author{T.A. Kamenskaya}
 
\affiliation{ 
P.N. Lebedev Physical Institute of the Russian Academy of Sciences. Leninsky Prospekt 53, 119991, Moscow, Russia
}%
\author{I.A. Eliseyev}%
\affiliation{ 
A.F. Ioffe Physico-Technical Institute. Politekhnicheskaya 26, 194021, St.Petersburg, Russia
}%

\author{V.Yu. Davydov}%
\affiliation{ 
A.F. Ioffe Physico-Technical Institute. Politekhnicheskaya 26, 194021, St.Petersburg, Russia
}%

\author{A.Y. Kuntsevich}
\affiliation{ 
P.N. Lebedev Physical Institute of the Russian Academy of Sciences. Leninsky Prospekt 53, 119991, Moscow, Russia
}%
 \email{alexkun@lebedev.ru}
\affiliation{%
HSE University, 101000, Moscow, Russia
}%

\date{\today}

\begin{abstract}
We have modified the metal-assisted transfer technique to obtain large-area few-layer flakes from transition metal dichalcogenides bulk crystals by introducing an initial stage - exfoliation of the bulk crystal onto an intermediate substrate, specifically a silicon wafer coated with polyvinyl alcohol. Following this, we thermally evaporate silver onto the sample and transfer the top layers of the crystal along with the silver layer to the target substrate. This technique allows the production of visually non-corrugated single- and few-layer flakes with high yield. A direct comparison of the micro-Raman and micro-photoluminescence spectra of flakes exfoliated using our method with the spectra of those exfoliated from scotch tape reveals differences in their properties. We identify signatures of deformations in the flakes exfoliated from the intermediate substrate, indicating the presence of static friction between the substrate and the flake. Our findings thus suggest a novel method to induce intrinsic deformation in 2D materials.
\end{abstract}

\maketitle


Mechanical manipulation of atomic-layer transition metal dichalcogenides (TMDC) and van der Waals heterostructures has recently led to the discovery of novel physics phenomena \cite{mak2014valley, Wang2018, kang2019nonlinear, Jin2021, lipatov2022} and the development of devices with unprecedented parameters, including detectors \cite{Kufer2015}, memory cells \cite{Chiang2021}, light-emission devices \cite{Ye2015}, gate-controlled superconductivity \cite{Wang2020, an2020}, spintronic components \cite{sierra2021}, and topological material elements \cite{mine2019laser}. This success has, in turn, spurred the development of wafer-scale TMDC technology \cite{zhang2022, marx2017, tang2020}, aimed at integrating 2D TMDC concepts into micro- and nano-electronics \cite{desai2016, chang2016large, dodda2022active, cao2022perspective}.

The realization of any novel physical concept or device prototype relies on precise laboratory technology. For 2D TMDCs, this process typically begins with the exfoliation of high-quality mono- and few-layer Van der Waals materials from single crystals. Mechanical exfoliation is particularly popular because it yields top-quality layers from almost any material. Once these layers are located, subsequent technological operations are performed, including heterostructure assembly, lithography, mesa-etching, and the patterning of contact and gate electrodes. Due to the fragility of 2D materials and the multitude of required operations, the likelihood of faults increases. Thus, there is a strong demand for reproducible and high-throughput methods to obtain individual layers.

The standard scotch-tape exfoliation of 2D materials from single crystals is known to produce contaminants and yield a low number of monolayers with limited area\cite{novoselov2005}. However, this method provides flakes of high structural quality. Many efforts have been made to increase the area and yield of mechanical exfoliation \cite{zhang2021,li2022,guo2023}.

In particular, it was found that substrate plasma-activation\cite{huang2015} and the coverage of the substrate with organic materials, such as polyvinyl alcohol, increase the yield of flakes\cite{huang2020} for scotch-tape exfoliation. However, the most significant increase in yield was achieved with metal-assisted exfoliation\cite{magda2015,desai2016,velicky2018,johnston2022,liu2020}. This method relies on the adhesion of metals to the surfaces of chalcogenides, which is stronger than the inter-layer adhesion. Consequently, when a metal film is detached from the surface of the layered crystal, the entire top layer is transferred. The metal is subsequently etched away, leaving the TMDC layers isolated.

Despite the large yield and sample size (up to several millimeters\cite{liu2020}), metal-assisted exfoliated flakes often suffer from cracks, folds, and corrugations due to the imperfections of the crystal's top surface and the transfer process itself. The top surface of a large crystal is rarely perfect. In this paper, we address this issue by exfoliating from thick flakes previously exfoliated onto a substrate covered with polyvinyl alcohol (PVA) instead of directly from the single crystal. These starting multilayer flakes are already single-crystalline, leading to the formation of uniform monolayer and few-layer flakes free of corrugations and with substantial lateral size (several hundred micrometers). We confirm the efficiency of this method through Raman and photoluminescence studies of the obtained mono and few-layers of transition metal dichalcogenides (TMDCs). Interestingly, we found that even after the transfer and removal of the metal, the few-layer flakes transferred onto Si/SiO$_2$ substrates are strained. This phenomenon can only be explained by significant static friction between the flake and the substrate, preventing the flake from relaxing. This spontaneous strain could potentially be a novel method for tuning the properties of TMDC layers in device prototypes assembled from these flakes. 

\begin{figure}[h]
  \centering
  \includegraphics[width=0.5\textwidth]{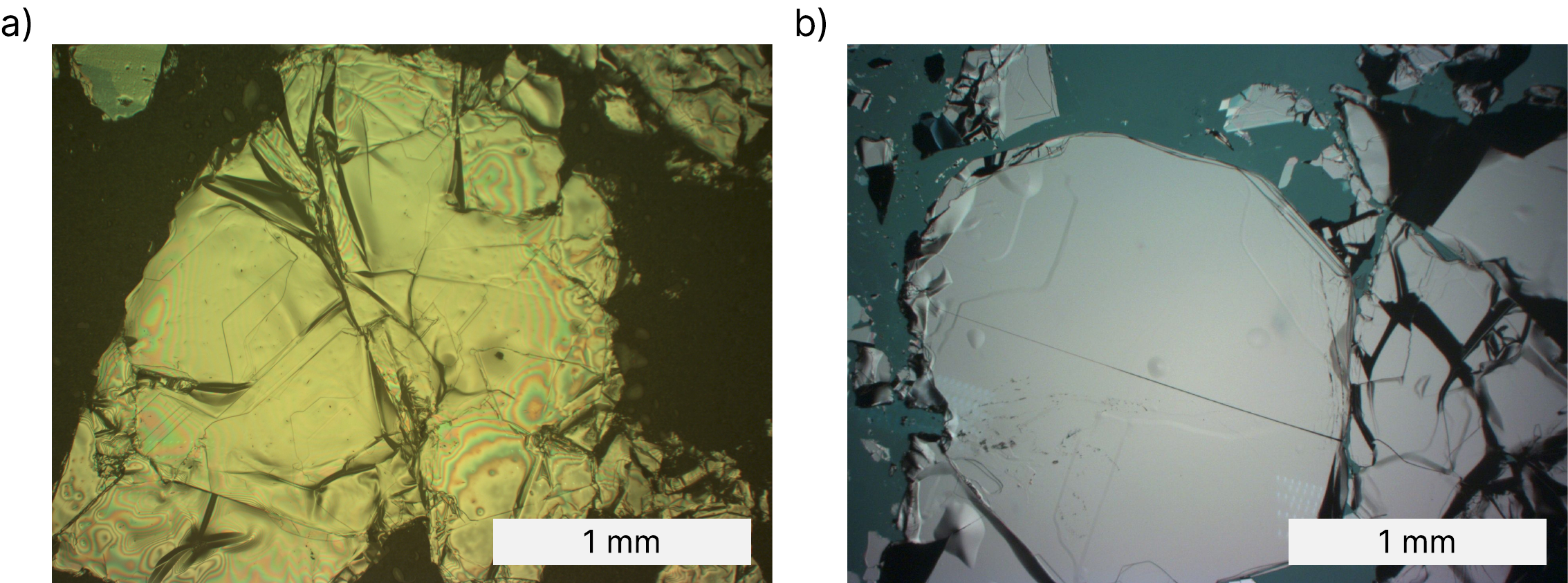}
  \caption{Mechanical exfoliation onto (a) scotch tape and (b) a substrate covered with a PVA thin film.}
  \label{fig: tape vs PVA}
\end{figure}

We began fabricating large-sized mono- and few-layer TMDCs using the method described in Ref.~\cite{desai2016}. Initially, we mechanically exfoliated natural molybdenite onto scotch tape. The flakes on the tape had large lateral sizes but exhibited uneven thickness, bubbles, and folds, as shown in Fig. \ref{fig: tape vs PVA}(a). These imperfections stem from the inevitable irregularities of the scotch tape and the surface of the parent crystal. To address these issues, we used a Si wafer coated with a polyvinyl alcohol (PVA) film as a temporary substrate instead of tape, increasing the yield, size, and homogeneity of the flakes\cite{huang2020}. {Additional comparative images of the bulk flakes on scotch tape and those on the Si substrate coated with PVA can be found in the Supplementary Information (Figure 1).}

\begin{figure}[h]
  \centering
  \includegraphics[width=0.5\textwidth]{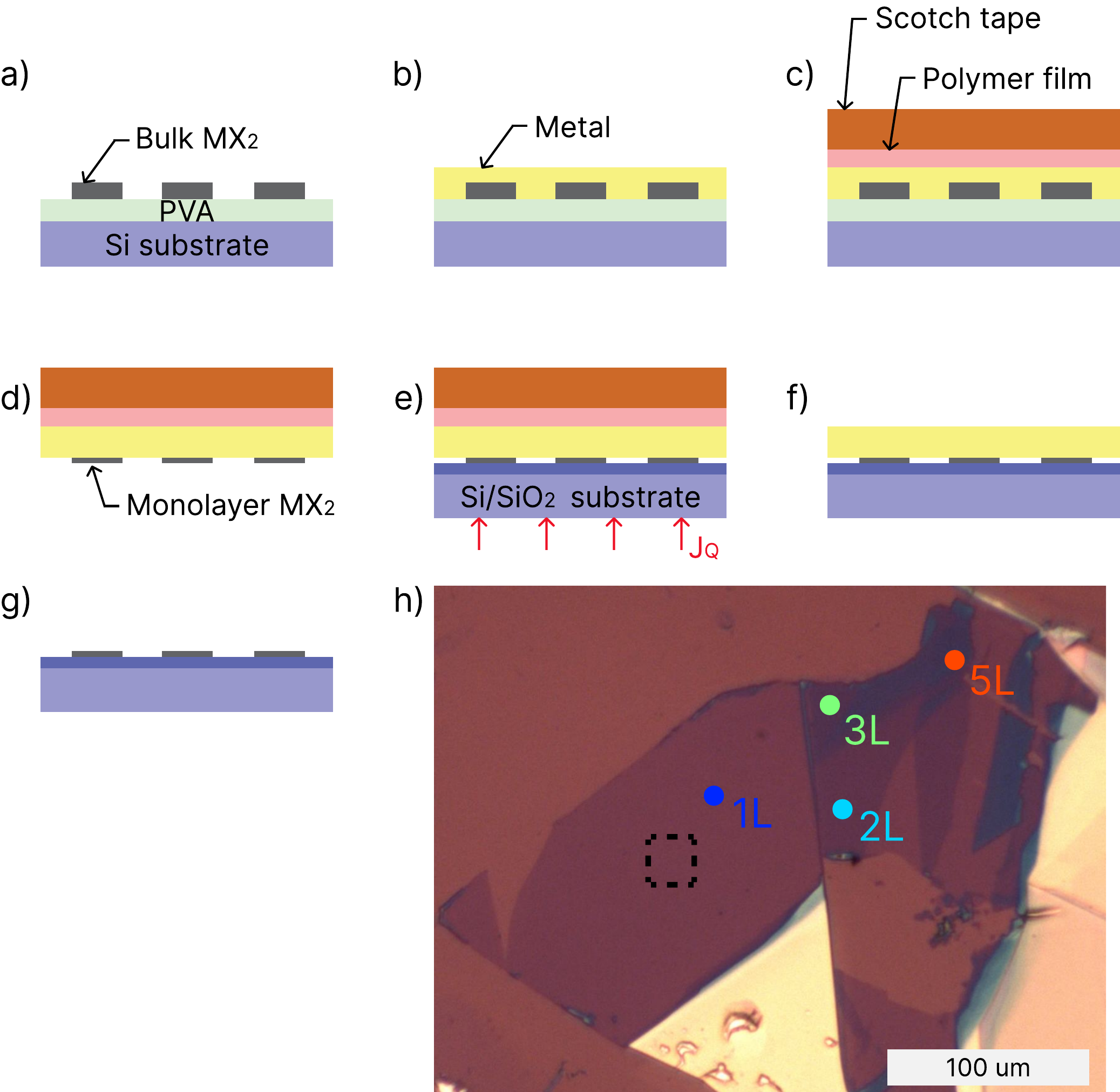}
  \caption{Stages of the exfoliation process. a) Mechanical exfoliation of a crystal onto a silicon substrate coated with a PVA film. b) Thermal evaporation of Ag and Al metal films. c) Spin-coating of a polymer film to serve as a buffer layer between the metal film and scotch tape. d) Peeling off the few-layer flakes from the temporary substrate. e-f) Transferring the metal film with flakes to a target substrate; peeling off the scotch tape and dissolving the polymer film. g) Resulting monolayer on the substrate after liquid etching of the metals and subsequent cleaning. h) Image of a MoS$_2$ sample showing 1L, 2L, 3L, and 5L areas. Colored circles indicate points where Raman spectra were collected. The black dashed rectangle highlights the region of PL mapping.}
  \label{fig: method}
\end{figure}

The PVA film was spun at 5000 rpm for 1 minute. An example of a large-sized flake with perfect crystallinity on PVA is shown in Fig. \ref{fig: tape vs PVA}(b). The process of bulk crystal exfoliation is illustrated schematically in Figure \ref{fig: method}(a).

After exfoliating the crystal onto the substrate, a thin silver film (50 nm) is e-beam evaporated atop, followed by \textit{in-situ} evaporation of aluminum (200 nm) to increase the rigidity of the film, as shown in Figure \ref{fig: method}(b). We used a {Plassys MEB 400} e-beam evaporation machine with multiple crucibles and a base vacuum of $5 \cdot 10^{-8}$ mbar. 

Instead of employing thermal release tape after metal deposition, we applied an additional methoxypropylacetate-based photoresist AZ-5214 film (Figure \ref{fig: method}(c)). The resist is spun at 3000 rpm for 1 minute. This organic layer softens the contact between the metal and tape, thereby preventing the layered crystal from cracking.

The topmost layers of the initial bulk flakes, now covered with metal, are peeled off onto the tape (https://aliexpress.ru/item/4000299853924.html) and transferred to the target substrate (Figure \ref{fig: method}(d)). The substrate, with the metal-coated flakes and the photoresist film, is then placed on a hot plate at 110$^\circ$C for a few minutes. During this heating process, the polymer solidifies and loses adhesion to the tape, enabling the removal of the scotch tape, as illustrated in Figure \ref{fig: method}(e-f). The substrate is then cleaned from the photoresist using acetone for approximately 30 seconds. To eliminate any organic residues, the sample undergoes treatment with oxygen plasma. The plasma etching parameters (100 W, 10$^{-1}$ mbar for 15 seconds) are carefully controlled to prevent etching of the both metal and the underlying monolayer. {The aluminum layer is crucial for protecting the silver film during oxygen plasma treatment from the oxidation \cite{pettersson1995preparation} }.

The aluminum film is etched using a photoresist developer for 40 seconds. Following this, the substrate is rinsed in distilled water to remove any residual developer. The silver film is then etched with a KI-I$_2$ wet etching solution for 30 seconds. Finally, the substrate, now containing the resulting monolayer, is cleaned in acetone for several minutes to remove any remaining residues, as illustrated in Figure \ref{fig: method}(g). From this point forward, we will refer to these samples as MA-PVA.

For comparison, we applied the same sequence of steps (illustrated in Figure \ref{fig: method}(b-g)) to crystals exfoliated on scotch tape, as shown in Figure \ref{fig: tape vs PVA}(a). These samples are referred to as MA-Scotch. To elucidate the role of the metal, we also performed exfoliation using standard mechanical exfoliation, employing only scotch tape and omitting the metal layer. These samples are designated as ME.

{We have successfully applied the MA-PVA method to other TMDC materials beyond MoS$_{2}$, including WSe$_{2}$ and WS$_{2}$, demonstrating its broader applicability. Images of these monolayers are shown in the Supplementary material. Additionally, we attempted to use the MA-PVA method on layered crystals such as Bi$_{2}$Te$_{3}$ and Sr$_{0.06}$Bi$_{2}$Se$_{3}$. These crystals is etched by the KI-I$_{2}$ solution used for silver removal. This limitation suggests that the MA-PVA method is particularly suitable for crystals that are resistant to such etching processes.}


Micro-Raman and micro-photoluminescence (PL) spectroscopy were employed for sample characterization. Raman and PL investigations were carried out using a Horiba LabRAM HREvo UV-VIS-NIR-Open spectrometer equipped with a confocal microscope. All measurements were performed in backscattering geometry at room temperature with continuous-wave (cw) excitation from the 532 nm laser line of a Nd:YAG laser (Laser Quantum Torus). For room-temperature Raman measurements, an Olympus MPLN100$\times$ objective lens (NA = 0.9) was used, which allowed us to obtain information from an area with a diameter of $\sim1$ $\mu$m. To prevent damage to the flakes, the incident laser power was limited to 200 $\mu$W. Raman spectra were obtained with a spectral resolution of 0.5 cm$^{-1}$ using a 1800 gr/mm grating. To suppress the Rayleigh scattering and obtain information from the ultralow frequency range (5-65 cm$^{-1}$) during Raman measurements, a set of Bragg filters (BragGrate) was used. Low-temperature PL measurements were carried out using a Linkam THMS600 stage, a long working distance objective lens (Leica PL FLUOTAR 50$\times$), and a 600 gr/mm grating. The PL spectra were obtained with $\sim2$ cm$^{-1}$ resolution. 

\begin{figure}[h]
  \centering
  \includegraphics[width=0.5\textwidth]{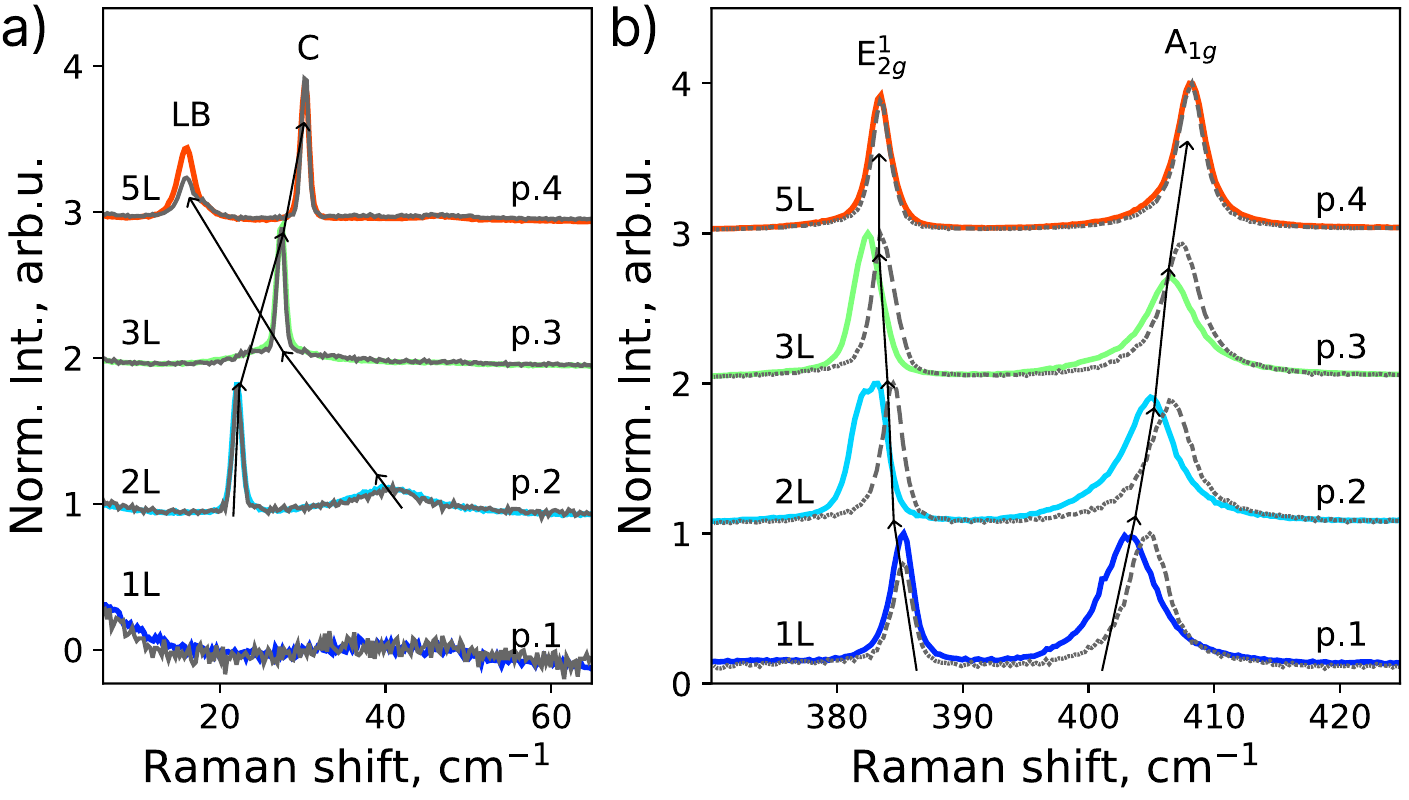}
  \caption{a) Low-frequency and b) high-frequency ranges of the Raman spectrum for few-layer flakes of MoS$_2$. The color solid lines represent the spectra of samples prepared using the MA-PVA method, while the gray dashed lines correspond to samples prepared with the MA-Scotch method.}
  \label{fig: Raman MoS2}
\end{figure}

Figure \ref{fig: Raman MoS2} presents the Raman spectra for MA-Scotch (dashed curves) and MA-PVA (color curves) samples, showing the low-frequency range in panel (a) and the high-frequency range in panel (b). 

In the low-frequency range, the Raman spectra for all three methods (ME, MA-Scotch, and MA-PVA) are nearly indistinguishable. This spectral range is used to determine the number of layers: no observable peaks are present for monolayers, while few-layer samples show two distinct peaks corresponding to the shear (C) mode and the layer-breathing (LB) mode \cite{zhang2013raman}. The frequency of the C-mode increases with layer thickness, whereas the frequency of the LB-mode decreases. 

The high-frequency range of the Raman spectrum, shown in Figure \ref{fig: Raman MoS2}(b), has two peaks corresponding to the E$_{2g}^{1}$ and A$_{1g}$ modes \cite{chakraborty2013layer}. The frequency of the A$_{1g}$ mode increases with the number of layers, while the E$_{2g}^{1}$ mode exhibits a decreasing trend. The peak positions, intensities, and linewidths in this range vary for 1L, 2L, and 3L flakes depending on the exfoliation method used. These variations are often attributed to factors such as doping or strain of the layers \cite{chakraborty2012symmetry, rice2013raman, wang2013raman, zhu2013strain, panasci2021}. The behaviour of the two high-frequency modes with changes in these factors is different: the position of the E$_{2g}^1$ mode is more sensitive to strain, whereas the A$_{1g}$ mode is more sensitive to doping \cite{chakraborty2012symmetry}. 

\begin{figure}[h]
  \centering
  \includegraphics[width=0.5\textwidth]{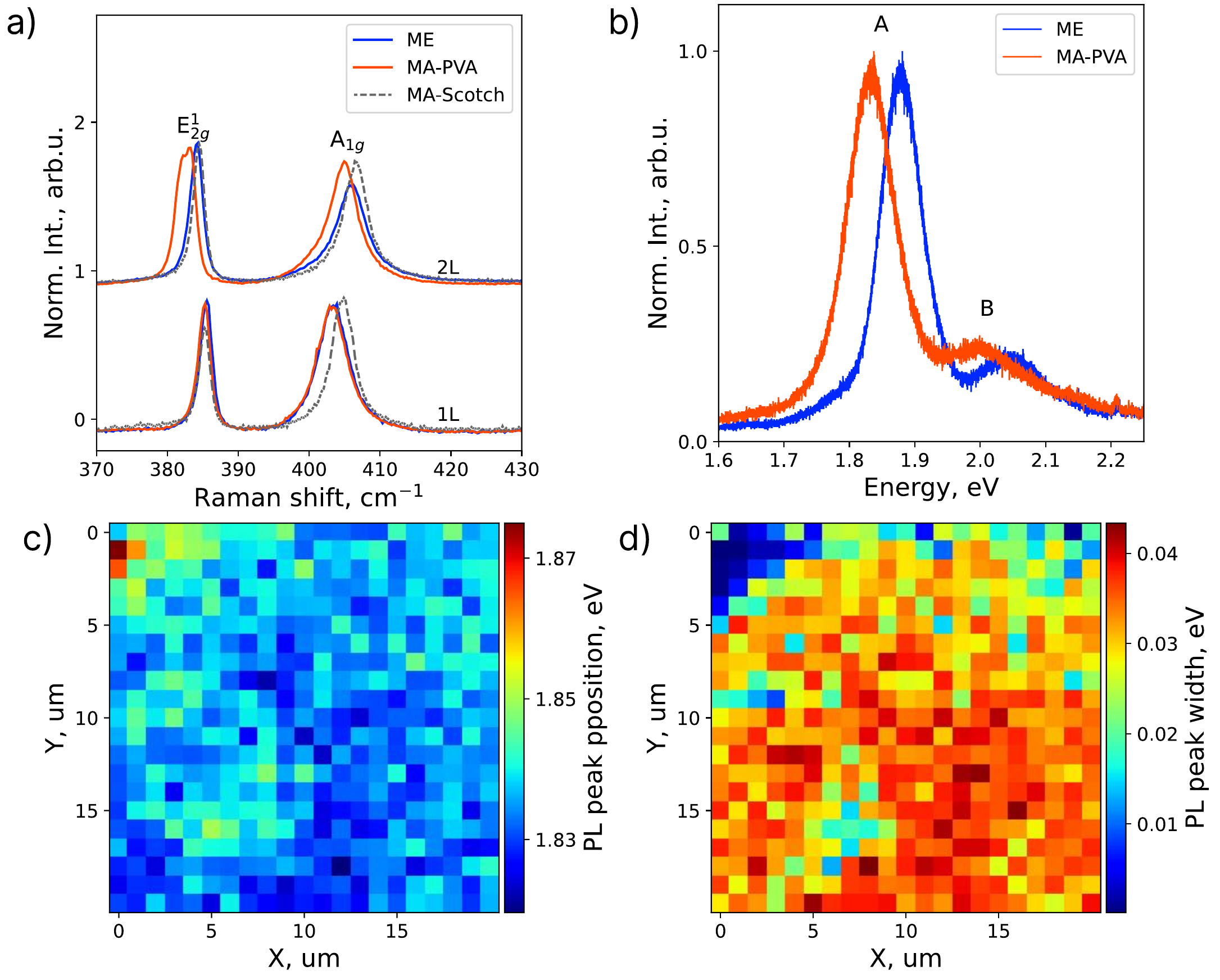}
  \caption{Comparison of a) Raman spectra for 1L and 2L MA-PVA (red) and ME (blue) MoS$_{2}$ flakes and b) PL spectra at 78K for MA-PVA (red) and ME (blue) monolayers. Maps of the MA-PVA A-exciton photoluminescence peak position  (c) and width (d) collected at 78 K across the rectangle area in Figure \ref{fig: method}(h).}
  \label{fig: mech vs LLE}
\end{figure}

The A$_{1g}$ Raman peak in the monolayer obtained using the MA-Scotch method is shifted by 2 cm$^{-1}$ compared to the monolayers prepared by the ME and MA-PVA methods. Additionally, the intensity ratio of the A$_{1g}$/E$_{2g}^{1}$ peaks is higher in the MA-Scotch samples. These observations suggest an increase in the p-doping of the MA-Scotch samples \cite{panasci2021,shi2013selective} compared to the MA-PVA ones. The Raman spectrum for the bilayer obtained by the MA-Scotch method also indicates p-doping (Figure \ref{fig: mech vs LLE}(a)). Electron- or hole-doping is a common phenomenon observed in metal-assisted exfoliated TMDCs \cite{panasci2021}. Notably, the E$_{2g}^{1}$ line position in monolayers remains constant (385.3 cm$^{-1}$, Figure \ref{fig: tape vs PVA}), pointing to the absence of strain in the MA-Scotch and MA-PVA samples \cite{panasci2021}. 

The frequencies, intensities, and widths of the E$_{2g}^{1}$ and A$_{1g}$ modes in the monolayers prepared by the ME and MA-PVA methods are similar, as shown in Figure \ref{fig: mech vs LLE}(b). 
In contrast, the bilayers produced by the MA-PVA method exhibit a significant redshift in both the E$_{2g}^{1}$ and A$_{1g}$ modes, along with splitting of the E$_{2g}^{1}$ mode. 
Such splitting is connected with lifting of the E$_{2g}^{1}$ mode degeneracy due to the presence of uniaxial strain \cite{wang2013raman}. Thus, the bilayers produced by MA-PVA are subject to uniaxial tensile strain. A similar effect was reported in Ref. \cite{velicky2020intricate} for the monolayers directly placed on Au and Pt. In Ref. \cite{panasci2021}, significant strain was observed in gold-assisted exfoliated MoS$_2$ layers on Ag. The effect of strain in the layers placed on different metals is due to the mismatch between the crystallographic structures of the TMDC and the metal \cite{velicky2020intricate}. Notably, in Ref. \cite{panasci2021}, the strain sign changed to compressive when the layers were transferred to Al$_2$O$_3$.


A further indication of strain is evident from the comparison of PL spectra for ME and MA-PVA MoS$_2$ monolayers, as shown in Figure \ref{fig: mech vs LLE}(b). The PL spectra for both layers exhibit two peaks: (i) a low-energy peak, corresponding to the direct A exciton (approximately 1.87 eV for the ME sample), and (ii) a higher-energy peak, corresponding to the direct B exciton (approximately 2.05 eV for the ME sample). Both excitons originate from the K point of the Brillouin zone due to spin-orbit splitting of the valence and conduction bands \cite{splendiani2010emerging}. For the MA-PVA monolayer, both peaks are significantly red-shifted by approximately 50 meV. This redshift is attributed to the presence of strain in the MA-PVA sample \cite{panasci2021}.

To characterize the uniformity of strain, we measured maps of the PL peak position (Figure \ref{fig: mech vs LLE}(c)) and width (Figure \ref{fig: mech vs LLE}(d)) of the A exciton peak across a 20$\times$20 $\mu$m$^2$ area of the MA-PVA monolayer. The maps reveal a smooth modulation of these properties, alongside a shot-noise-like component with variations in the position of the A peak up to 50 meV, indicating nanoscale emitters and possibly defects.  

Notably, the spectral position and linewidth show a correlation: higher energy corresponds to narrower linewidths. This observation aligns with the deformation model, where non-deformed regions are structurally more perfect and exhibit narrower spectral lines.
The average PL peak half-width of the MA-PVA monolayer is greater than the typical linewidth observed for the ME monolayers. This increased linewidth may be attributed to the presence of trions, which are indicative of elevated doping levels in metal-exfoliated samples \cite{petrini2024}.




In our study, MoS$_2$ flakes on the same substrate, obtained using different exfoliation methods, exhibit either tensile strain (MA-PVA) or no strain (ME, MA-Scotch). This indicates that the MA-PVA flakes do not fully relax after being transferred onto the substrate and after the silver has been etched off. This suggests the presence of finite static friction that prevents the sample from achieving a relaxed state. The static friction likely arises from local defects that disrupt the translation symmetry of the system. These defects are distributed non-uniformly, leading to variations in optical properties across the sample area. Identifying and characterizing the amount and spatial distribution of these defects is challenging. A potential mechanism for their formation is the damage to the TMDC layer during metal deposition \cite{heyl2023}, which can induce strain and doping.

In contrast to the MA-PVA samples, which exhibit a perfectly flat surface and a strong adhesion to the substrate, the MA-Scotch samples show numerous folds, scratches, and bubbles. Due to their weaker bond to the substrate, the strain in MA-Scotch samples generally relaxes more effectively.

Another indirect indication of the strong friction between the substrate and the flake is the inability to detach the metal-exfoliated flakes from the substrate using the dry transfer method \cite{pizzocchero2016hot} (which we tested). In contrast, flakes exfoliated using scotch tape can be picked up with relative ease.

Although MoS$_2$ is recognized for its solid lubricant properties, nonzero static (or dry) friction in TMDC flakes has been experimentally reported \cite{gan2019} using scanning probe microscopy. We anticipate that similar experiments with metal-exfoliated layers will reveal the complexities and properties of mechanically strained layers.


Mechanical deformation of flakes is a well-established phenomenon with broad implications. Since strain influences a wide range of properties \cite{peng2020strain}, it is crucial to consider this deformation when assembling devices from mono- or few-layer flakes. For instance, the recently investigated phenomenon of Moir? ferroelectricity \cite{wang2023towards} requires precise alignment of two flakes and is highly sensitive to strain. Additionally, exploring methods to return flakes to a non-relaxed state using techniques such as thermal cycling, laser pulses, electron beams \cite{du2022strain}, or scanning probe microscopy is essential for fully understanding and controlling these effects.

The MA-PVA method also offers practical advantages due to the low cost and availability of its consumables. Notably, silver is approximately 40 times cheaper than gold, which is commonly used in metal-assisted exfoliation \cite{heyl2023}. Additionally, instead of relying on costly thermal release tape, our method employs inexpensive photoresist in combination with widely available scotch tape.

In conclusion, we have developed an enhanced metal-assisted exfoliation technique for two-dimensional materials. This method involves first exfoliating large, structurally pristine flakes onto a temporary substrate coated with polyvinyl alcohol, followed by metal (silver) deposition. The top one-to-few layers are then peeled off and transferred to an oxidized silicon surface. Our observations reveal that the flakes are initially strained due to the metal, and this strain persists even after the metal is etched away. This indicates significant adhesion and dry friction between the layers and the substrate. Strained flakes offer valuable insights for both fundamental physics and practical applications, providing new opportunities in the field of 2D materials.

\section{Acknowledgements.}
The authors would like to express their gratitude to A.I. Duleba for conducting the initial experiments and for his dedication in finalizing the paper, and to M.V. Pugachev for his valuable assistance. This work was supported by the Russian Science Foundation under Grant No. 23-12-00340. A.Y.K. acknowledges support from the Basic Research Program of the Higher School of Economics (HSE). Sample fabrication was conducted at the Shared Facility Center of the P.N. Lebedev Physical Institute, Raman and PL measurements were performed at the Ioffe Institute. I.A.E. and V.Y.D. acknowledge partial support within the framework of
the state assignment of the Ministry of Science and Higher Education
(FFUG-2024-0018).

\section*{Data Availability Statement}
Data available on request from the authors.

\appendix

\nocite{*}
\bibliography{arxiv}

\end{document}